\documentclass[12pt]{amsart}
\input epsf.tex
\usepackage{latexsym}
\usepackage{amsmath,amsthm,amsfonts,amssymb}
\usepackage{epsfig}

\newtheorem{pro}{Proposition}[section]
 \newtheorem{thm}[pro]{Theorem}
 \newtheorem{lem}[pro]{Lemma}

 \newtheorem{cor}[pro]{Corollary}

 \def\O{{\rm O}}

\def\Per{{\rm Per}}

 \def\R{{\mathbb R}}
 \def\chix{{\raise.5ex\hbox{$\chi$}}}
 
\def\Z{{\mathbb Z}}

\def\tM{{\widetilde M}}

\def\tX{{\widetilde X}}

\def\f{{f}}

\def\ep{{\varepsilon}}

\def\Sym(#1){\Sigma_{#1}}
\def\limmu_#1{\mu^{(#1)}}
\def\cvxmu_#1{\lambda_{#1}}
\def\infrac#1#2{(#1)/(#2)}
\def\Seq#1{\langle #1 \rangle}
\def\uP{\widehat P}
\def\II#1{{\bf 1}_{\{#1\}}}

\begin{document}
\title{A Solidification Phenomenon in Random Packings}
\author{L. Bowen, R. Lyons, C. Radin, \and P. Winkler}
\address{Department of Mathematics\\ Indiana University\\ Bloomington,
  IN 47405}
\email{lpbowen@indiana.edu}
\address{Department of Mathematics\\ Indiana University\\ Bloomington,
  IN 47405}
\email{rdlyons@indiana.edu}
\address{Department of Mathematics\\ University of Texas\\ Austin, TX 78712}
\email{radin@math.utexas.edu}
\address{Department of Mathematics\\ Dartmouth College\\ Hanover, NH 03755}
\email{peter.winkler@dartmouth.edu}
\thanks{Research was partially supported by the National Science
Foundation under grants DMS-0406017 and DMS-0352999. The authors also 
thank the Banff International Research Station for support at a workshop
where this research was begun.}
\date{}

\begin{abstract}
We prove that uniformly random packings of copies of a certain simply-connected
figure in the plane exhibit
global connectedness at all sufficiently high densities,
but not at low densities.
\end{abstract}
\maketitle


\section{Introduction}

\begin{figure}[htp]
\epsfxsize=2.8truein
$$\epsfbox{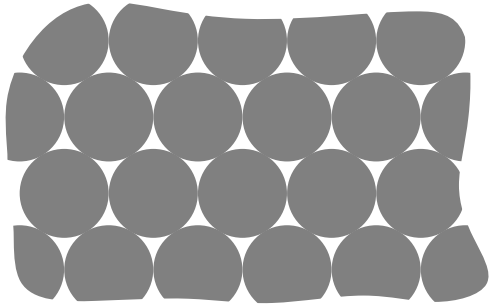}$$
\caption{The densest packing of unit disks in the plane}\label{fig:disks}
\end{figure}

The densest way to cover a large area with non-overlapping unit disks
is as in Figure~\ref{fig:disks}.

\noindent
A {\bf packing} is a
collection of congruent copies of a subset with pairwise disjoint interiors.
See \cite{Fej}
for a proof that the above packing is indeed densest possible for unit disks.

It is an old unsolved problem to understand whether densest packings
of spheres, simplices or other shapes,
in a Euclidean or hyperbolic space of any dimension,
exhibit
crystallographic symmetry.  For instance, this is the spirit of Hilbert's
eighteenth problem; see \cite{Fej, Rad} for background. 

Using physics models of two- and three-dimensional matter as a guide, we are
tempted to try to gain insight about densest packings by considering
packings at 
densities below the maximum. (For an example concerning
spheres in $\R^3$, see \cite{KRS}.) In effect, we are emphasizing
not so much that densest packings and sparse packings differ by their
{\em symmetry}, as that they differ in some fundamental geometric
fashion. Indeed, it is
commonly suggested in the physics literature (see for instance
\cite{AH}) that 2-dimensional models of 
matter do not exhibit crystallographic symmetry, and it is sometimes
said by mathematicians that in high dimensional Euclidean space,
densest packings of spheres may not have crystallographic symmetry. So
perhaps it is appropriate to reexamine the precise manner in which
densest packings differ fundamentally from sparse packings, and to use
packings at less than optimum density as a guide.

The density of the unit disk packing of Figure~\ref{fig:disks} is $\widetilde d:=
{\pi/ \sqrt{12}}\approx 0.91$; packings of density
0.89 (say) can be obtained by shrinking the disks slightly, shaking
the whole collection, and then
expanding the whole picture to recover unit-size disks.  
One might ask whether by doing this one preserves long-range correlations.
Experiments and discrete models indicate that such long-range
correlations
are indeed preserved, but this has never been proved.
See \cite{BLRW} for background.

In this work we prove the conjectured behavior, not for disks but for deformed
disks, copies of a ``zipper'' tile designed specifically for the
purpose; see Figure \ref{fig:tile}.
This tile can cover the plane completely, in which case the packing has
density 1, and is completely connected in any sense.
What we show is that even at somewhat lower densities, the uniform random
packing still has rich structure; in particular it has a form of
connectedness associated with site percolation \cite{G}.
What this means for packing large but finite boxes (with torus boundary
conditions) is that the necessary gross irregularities of most packings at such
high densities occur, but not in a way to disconnect the packings.
Although we define ``uniform random packing" of the plane by limits of
measures on packings of finite boxes, the key to our proof is to examine
isometry-invariant probability measures on packings of the whole plane and
to show that the ones that maximize ``degrees of freedom per tile"
are unique for high densities.

In physics one is interested in change of behavior as density
changes.
We show that at high density there is a nonzero probability of
an infinite linked component, and that this probability is zero at
low density.
Thus, there are
different ``phases'' of the packings \cite{BLRW}.
(This is the same phase transition seen in continuum percolation, where one
looks at overlapping disks with random independent centers, but our methods
are quite different. Indeed, no such result is known for packings of disks.)

Although we believe such a result also holds for packings of
disks or of spheres --- pairs of which would be called ``linked''
if sufficiently close --- we are able to prove the result only for our tiles,
which are shaped to allow three well-defined levels of pairwise
separation. It is generally understood that crystalline behavior
is not seen in two dimensions, so the form of connectedness we use
may be useful in understanding the role of geometry in Hilbert's problem.

\section{Description of the tile}

\begin{figure}[htp]
\epsfxsize=2.7truein
$$\epsfbox{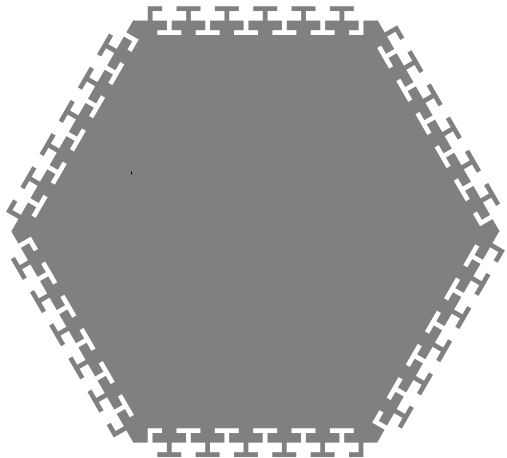}$$
\caption{The zipper tile}\label{fig:tile}
\end{figure}

We consider packings by a deformed disk denoted by $t$,
referred to as ``the tile'' and depicted in Figure \ref{fig:tile}. In this
section, we define it precisely.

Let $H$ be a regular hexagon of area 1. Let $r$ be the radius of the in-circle of $H$.
Let $D$ be a disk concentric with the in-circle and of radius $r+\rho$,
where $0<\rho\ll 1$
is a number we shall choose more precisely later. We shall construct the shape $t$  by modifying $H$ as follows; $D$ will be called
the {\bf shadow disk} of $t$.

\begin{figure}[htp]
\epsfxsize=3.7truein
$$\epsfbox{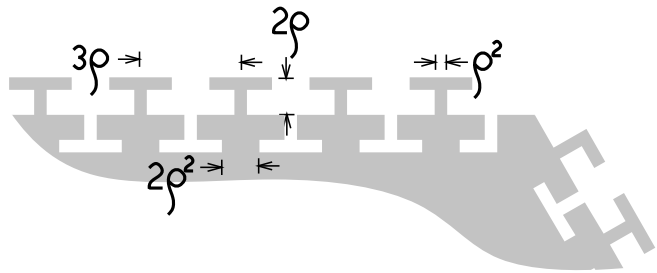}$$
\caption{Close-up view of the fringe}\label{fig:measure}
\end{figure}

\begin{figure}[htp]
\epsfxsize=3.5truein
$$\epsfbox{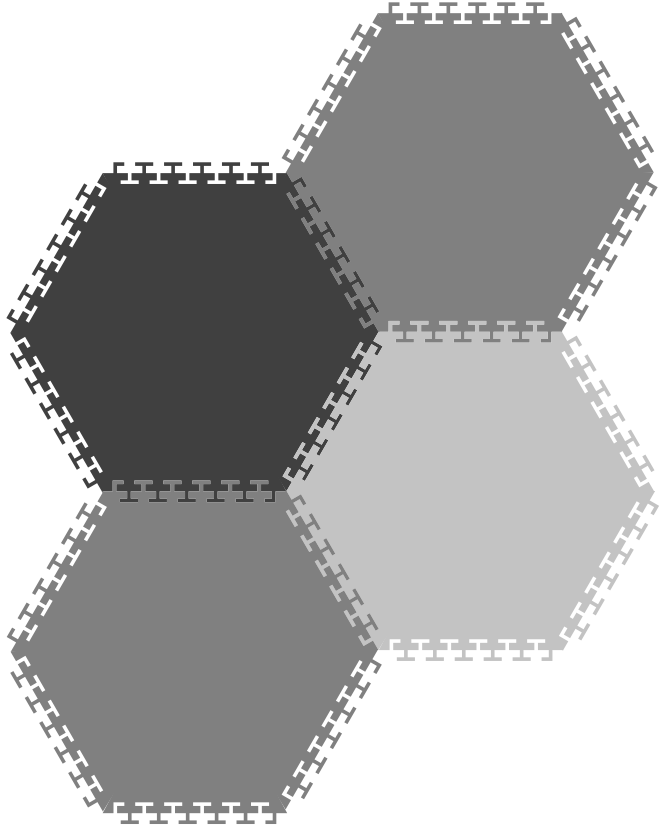}$$
\caption{Tight-linked tiles}\label{fig:tight}
\end{figure}

\begin{figure}[htp]
\epsfxsize=3.5truein
$$\epsfbox{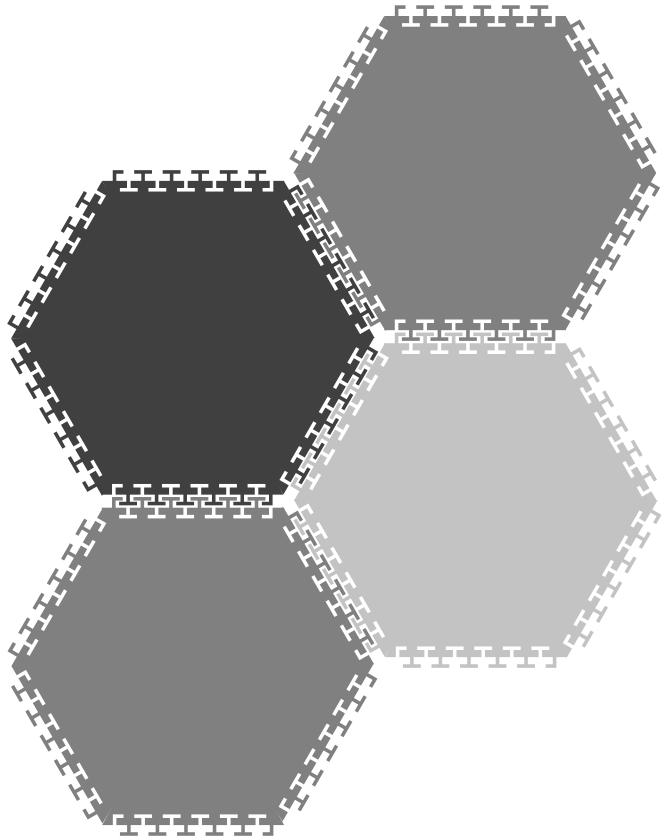}$$
\caption{Loose-linked tiles}\label{fig:loose}
\end{figure}

As shown in Figure \ref{fig:tile}, the tile $t$ equals $H$ with each side modified
by a ``fringe'' and each corner modified by a hook and inlet, where a
hook is about
half an element of the
fringe. As shown in Figure \ref{fig:measure}, the fringe height
is $2\rho$.  The elements of the fringe have two different size ``necks'', one of size
$\rho^2$ and one of size $2\rho^2$, allowing neighboring tiles to be linked in either
of two well-defined modes, ``tight linked'' and ``loose linked'', the
former illustrated
in Figure \ref{fig:tight} and  the latter illustrated
in Figure \ref{fig:loose}.  
We say that two tiles $t$ are
{\bf linked} (tightly or loosely) if when one is held fixed, the
other can be moved continuously only by a bounded amount (without
overlapping the first).  
A {\bf tight} link is one that permits no movement of one tile while fixing
the other, while a link that is not tight is called {\bf loose}.
A key feature of our model is that when two tiles are tightly
linked, any motion of one would necessitate a corresponding motion of the
other.  As we shall explain, the uniform
probability distribution on packings of the plane at given density is a
limit of such a distribution on packings of larger and larger tori.  In our
model,
these distributions on packings of finite tori are concentrated on packings with
the maximal number of degrees of freedom, and therefore intuitively the
fewest possible number of 
tiles bound by tight links.  This gives us useful control on the packings in the support
of our distributions.

A tile is called {\bf fully linked on one side} if it is linked with
another tile on that side in such a way that either
they are tight linked and the line joining their centers goes through the
midpoint of the sides of the corresponding hexagons; or
the tiles are not tight linked but can be moved continuously so that their
shadow disks touch each other.
A tile is {\bf fully linked} if it is fully linked on all sides.
We note that the fully tight-linked packing (Figure~\ref{fig:tight}) corresponds to a
tiling by the original hexagon and has density 1, and that the tile has area 1 by construction.

\section{Statement of Results}

To state our results we need some notation. Let $X$ be the space of
all packings of the plane by the tile $t$. Given a compact subset $K$ of the
plane and two packings of the plane, we consider the distance between the
two packings with respect to $K$ to be the Hausdorff distance between the
unions of the tiles in the respective packings intersected with $K$.
Then $X$ is endowed with the
topology of Hausdorff convergence on compact subsets; $X$ is compact.
Intuitively, two
packings are close in $X$ if they are close in the Hausdorff sense in
a large ball centered at the origin. We shall define a probability
measure on $X$ that is ``uniform'' on the set of all packings of a
fixed density. For this, we shall need the space $X_n$ of all packings
by the tile of the $n \times n$ torus $\R^2/(n\Z)^2$.

For any integer $m$, let $X_{n,m} \subset X_n$ consist of those packings which
contain exactly $m$ tiles ($X_{n,m}$ is empty if $m$ is large
enough). To each tile, we assign the set of $6$ unit vectors based on
its center and pointing to the center of each of its edges. Through
this assignment, we can view $X_{n,m}$ as a subset of $T_n^m/\Sym(m)$,
where $T_n$ is the unit tangent bundle of the $n\times n$ torus modulo
a $2\pi/6$ rotation and the symmetric group $\Sym(m)$ acts by permuting
the factors.

When $m/n^2$ is small, $X_{n,m}$ is $(3m)$-dimensional. However, when $m/n^2$ is
sufficiently large, the dimension of $X_{n,m}$ inside $T_n^m/\Sym(m)$
is less than $3m$. This is because at least two tiles in any
packing of $X_{n,m}$ will have to be tightly linked, so that it is
impossible to move one continuously without moving the other. Thus it is
useful to decompose
$X_{n,m}$ into a (finite) disjoint union of sets $X_{n,m,k}$ of
packings containing exactly $k$ tight links. Generically, the
dimension of $X_{n,m,k}$ is $3(m-k)$. The dimension can be strictly
less than this if the packings are jammed in the sense of \cite{DTSC},
although this fact will not be important for us. The {\bf top dimension} of
$X_{n, m}$ means the maximum dimension of all $X_{n,m,k}$. Let
$\mu_{n,m}$ be the probability measure on $X_{n,m}$ obtained by
normalizing the Hausdorff measure on $X_{n,m}$ in the top dimension of
$X_{n,m}$ with respect to the natural
metric inherited from $T_n^m/\Sym(m)$. We interpret $\mu_{n,m}$ as
being a uniform measure. The fact that $\mu_{n,m}$ is supported on
those packings with the fewest number of tight links will be crucial in
the analysis to follow.

Let $\tX_{n}$ be the space of all $(n\times n)$-periodic packings of the
plane. In other words, $\tX_{n}$ consists of those packings that are
preserved under
translations by $n\Z \times n \Z$. Under the quotient map, this space
is naturally identified with $X_{n}$. Therefore, we can view the
measures $\mu_{n,m}$ as living on $\tX_n \subset X$.

For a fixed density $d \in [0,1]$, let $\limmu_d$ be any measure obtained
as the weak* limit of measures of the form $\mu_{n,m}$ such that $n
\to \infty$ and $m/n^2 \to d$. (Note that $m/n^2$ is the density of every
packing in the support of $\mu_{n,m}$ and $d$ is the average density
of a packing chosen with respect to $\limmu_d$: see Lemma \ref{lem:cont}.)
A priori, $\limmu_d$ may not
be unique, although we shall prove that it is for large enough $d$.

As we have seen, two tiles $t$ can be
linked.  Therefore it makes sense to speak of a {\bf linked
component} of a packing; it is a maximal sub-packing such that for every two
tiles $t,t'$ in it, there is a sequence $t=t_1, t_2,\ldots,t_n=t'$ such
that $t_i$ is linked to $t_{i+1}$ ($i=1,\ldots,n-1$). A {\bf tight-linked component} is defined similarly, but we require $t_i$ to be tightly-linked to $t_{i+1}$.

We say that a measure on the space $X$ of packings is invariant if it preserved under the full isometry group of the plane. All the measures we consider are probability measures unless stated otherwise. 

Let $\lambda_0$ be the unique invariant measure on tilings (packings that
cover $\R^2$)
by our tile.  
Let $\lambda_1$ be the unique invariant measure
on packings by $t$ such that all tiles are fully loose linked, are as close
as possible to each other, and the packing has hexagonal symmetry. 
Write $\cvxmu_s:=s \lambda_1 + (1-s)\lambda_0$.

 Our main results are the following:

\begin{thm}\label{thm:main1} There exists $0<d_1 < 1$ such that if $d
  \ge d_1$, $\limmu_d$ is unique and equals $\cvxmu_s$,
  where $s:=\infrac{1-d}{ 1-d_1}$.
\end{thm}

\begin{cor}
  The $\limmu_d$-probability that the origin is inside a tile belonging to
  an infinite linked component is nonzero for $d \ge d_1$.
\end{cor}

\begin{pro}\label{thm:low1} For some $d_2>0$, the probability (with
  respect to any $\limmu_d$ for any $d < d_2$) that the origin
is inside an infinite linked component is zero.
\end{pro}

\section{Tile properties}

\begin{lem}
  For small $\rho$, 
  if tiles $t_1$ and $t_2$ are not tightly linked and do not overlap, then the
  distance between their centers is at least $2r + 2\rho$.
\end{lem}
\begin{proof}
  Consider the line segment from the center of $t_1$ to the center of
  $t_2$. If this segment traverses near a corner of $t_1$ or $t_2$,
  then it must be longer than $2r+2\rho$ for small enough $\rho$.
  Supposes it crosses a fringe of $t_1$ and of $t_2$. If the tiles
  are not linked, then the claim is obvious. If they are linked, then to
  minimize the distance, it must be that their fringes match up (so
  they are fully linked on one side). Thus the closest they can come
  is if the two are pushed flat up against each other so that their shadow
  disks touch. In this case, the distance between the centers is exactly
  $2r+\rho$.
\end{proof}

We shall say that two tiles are {\bf densely loose linked} if they are
loose linked and their shadow disks touch.
There is a unique invariant measure on maximally
dense packings by congruent disks \cite{BHRS}.
Hence the probability measure $\lambda_1$ that we defined earlier is the
unique invariant measure on packings by $t$ such that all tiles are fully
and densely loose linked. Let $d_1$ be the density of such a packing.

Given a tile $t$ in a packing $P$, we denote by $V(t)$ the Voronoi cell of
the center of $t$ with respect to the centers of the other tiles; that
is, $V(t)$ is the open set of points closer to the center of $t$ than to the
center of any other tile.
  We denote the area of a region $A$ of the plane by $|A|$.

\begin{lem}\label{lem:lambda1}
  The following holds for small enough $\rho>0$. For any packing $P$,
  if $t \in P$ is a tile that has no tight links, then the area of
  $V(t)$ is least $1/d_1$. Moreover, equality holds iff the configuration of
  tiles determining $V(t)$ is congruent to a corresponding
  configuration of a packing in the support of $\lambda_1$.
\end{lem}

\begin{proof}
  For a tile $t$, let $H(t)$ denote the hexagon from which $t$ is created.
  For $x > 0$, let $H_x(t)$ denote the homothetic copy
  $\frac{r+x}{r}H(t)$ about the center of $H(t)$.

  Suppose $t$ is a tile of $P$ without any tight links.  Consider
  the rays $R_1,\dots,R_6$ from the center of the hexagon $H(t)$ through
  each of its 6 vertices. These rays divide the plane into 6 sectors,
  $S_1,\dots,S_6$. 
  
   By construction, if $t$ and $t_1$ are
  loose linked, then $|H_{\rho}(t) \cap H_\rho(t_1)| = \O(\rho^2)$:
  The hexagon interiors do not intersect if they are parallel, while
  if they are not parallel, they can intersect only very slightly at
  a corner. The openings at
  which a corner can enter have area $\O(\rho^2)$ as $\rho\to 0$.

  Thus, we have proved that whenever $t$ is
  loose linked in the sector $S_i$, then $|V(t) \cap S_i| \ge
  |H_\rho(t)|/6 - \delta_1$, with $\delta_1 =\O(\rho^2)$ as $\rho\to 0$.
  
  Similarly, if $t$ and $t_2$ are not linked at all, then $|H_{2\rho}(t)
  \cap H_{2\rho}(t_2)| = \O(\rho^2)$:  Again, their
  interiors do not intersect if they are parallel, while if they
  are not parallel, they can intersect only very slightly at a
  corner. So 
  there exists $\delta_2>0$ such that whenever $t$
  is not linked in the sector $S_i$, we have $|V(t)\cap S_i| \ge |H_{2\rho}(t)|/6 -
  \delta_2$, with $\delta_2 =\O(\rho^2)$ as $\rho\to 0$.  
  
  Therefore, if $t$
  has no tight links but is not fully linked, then
$$
|V(t)| \ge j\Big( \frac{|H_\rho(t)|}{6} -\delta_1\Big) +
(6-j)\Big(\frac{|H_{2\rho}(t)|}{6}-\delta_2\Big)
$$
for some $j$ with $0 \le j \le 5$. Given that $\delta_1,\delta_2$ are
of order $\rho^2$ while $|H_{2\rho}(t)|-|H_{\rho}(t)|$ is of order
$\rho$, for $\rho$ small enough we may conclude that $|V(t)| >
|H_\rho(t)|$ in this case. 

On the other hand, the geometry of a tile is such that for small $\rho$, if
$t_1$ and $t_2$ are two tiles loose linked to $t$, then $t_1$ cannot be
tight linked to $t_2$. Now suppose that $t$ is fully loose linked. Then
the Voronoi cell of the center of $t$ is determined by six tiles
$t_1,\dots,t_6$ all loose linked to $t$ and all with the property that
their shadow disks $D,D_1,\ldots,D_6$ do not overlap (by the previous lemma). It follows 
\cite{Fej2} that $|V(t)| \ge |H_\rho(t)|$, with equality iff each of the disks
$D_1,\dots,D_6$ touches $D$. But there is only one way in which this can
occur (up to isometry). So $V(t)=H_\rho(t)$ in this case. This implies that the configuration
$t,t_1,\dots,t_6$ is congruent to a corresponding configuration of a
packing in the support of $\lambda_1$.
\end{proof}

It is easy to see that given $\rho > 0$, there exists $\ep>0$ such that for any finite
component $c$ of tight-linked tiles in any packing, the union $V_c$ of the Voronoi cells
of the centers of the tiles of $c$ has area at least $j_c +\ep \Per_c$.
Here $j_c$ is the number of tiles in $c$ and $\Per_c$ is the perimeter
of the union of hexagons corresponding to $c$. Let $\ep$ be the largest
such constant. Let $\delta>0$ be such that the area of the Voronoi cell
in the fully densely loose-linked packing equals $1+\ep \Per_1 + \delta$,
where $\Per_1$ is the perimeter of the hexagon of a single tile. Since
$\ep = \rho + \O(\rho^2)$ and $\delta = \O(\rho^2)$, we have:

\begin{lem}\label{lem:voronoi}
For sufficiently small $\rho$, there are $\ep, \delta > 0$ such that for
any finite tight-linked component $c$,
\begin{eqnarray*}
d_1 &=&\frac{1}{1+ \Per_1 \ep + \delta}\,,\\
|V_c| &\ge& j_c+ \ep \Per_c\,, \\
\noalign{\hbox{and}}
\delta &\le& \ep/100\,.
\end{eqnarray*}
\end{lem}

\section{High Density}
Recall that $X$ is the compact space of all packings of the plane by the tile
(with the topology of Hausdorff-metric convergence on compact
subsets). Let ${\widetilde M}$ be the space of isometry-invariant Borel probability
measures on $X$. For any $\mu \in {\widetilde M}$, we denote by $|\mu|:=\mu(A_0)$
the {\bf density} of $\mu$, where $A_0$ is the set of all packings $P \in X$,
one of whose tiles contains the origin.  Since a tile is the closure of its interior,
$A_0$ is a closed set.
\begin{lem}\label{lem:cont}
If $\mu_i \in {\widetilde M}$ converges to $\mu$ in the weak* topology, then
$|\mu_i|$ converges to $|\mu|$.
\end{lem}
\begin{proof}
Let $\uP$ denote the union of tiles in a packing, $P$.
For any invariant probability measure $\nu$ and any $z \in \R^2$, we have
$$
|\nu| = \int \II{0 \in \uP} \,d\nu(P)
=
\int \II{z \in \uP} \,d\nu(P)
\,.
$$
Integrating over $z$ in a unit-area disk, $D$, with respect to Lebesgue
measure and using
Fubini's theorem gives the identity $|\nu| = \int |\uP \cap D| \,d\nu(P)$.
Since the function $P \mapsto |\uP \cap D|$ is continuous on $X$, the
lemma follows.
\end{proof}

Recall that $\lambda_0$ is the unique invariant measure on tilings 
by our tile, so that $|\lambda_0|=1$.  Recalling that $d_1$ is the density
of a fully densely
loose-linked tiling, fix a density $d$ with $d_1 \le d \le 1$. Let $\mu_N$
be the uniform measure on configuration of tiles at density $d_N$ in an $N \times N$
torus, where $d_N \to d$ as $N \to \infty$. To prove theorem
\ref{thm:main1}, we shall show that the weak* limit of $\mu_N$ exists and
equals $\cvxmu_s$, where $s:=\infrac{1-d}{ 1-d_1}$.


We shall use several lemmas that depend on the following
notation.
 Given a packing $P \in X$, let
\begin{itemize}
\item $t_P$ be the tile of $P$ such that the origin belongs to $V(t_P)$
(this exists as long as the origin is not on the boundary of a Voronoi
cell),
\item $K_P$ be the tight-linked component containing $t_P$,
\item $j_P$ be the number of tiles in $K_P$, and
\item $\f(P) := 3/j_P$ if $j_P$ is finite and $t_P$
  contains the origin, and 0 otherwise.
\end{itemize}
Thus $\f(P)$, in a sense, measures the number of degrees of freedom per tile near
the origin.

\begin{lem}\label{lem1} If $\nu$ is any measure in $\tM$, then $\int f\,d\nu(\f)
  \le 3|\nu|$, with equality iff $t_P$ has no tight links for $\nu$-almost
every packing $P$.
\end{lem}
\noindent
The proof is immediate.

\begin{lem}\label{lem:fcontinuity} If a sequence $\Seq{\nu_n} \subset \tM$ 
converges to $\nu$ in the weak* topology, then $\int f\,d\nu_n$
converges to $\int f\,d\nu$.
\end{lem}

\noindent
Lemma \ref{lem:fcontinuity} is proven in a manner similar to Lemma \ref{lem:cont}.

Given a finite tight-linked component $c$, let the congruence class of $c$
be $C$ and let
$X_C \subset X$ be the space of
all packings $P$ for which $t_P$ exists and
$K_P$ is in $C$. Let $X''$ be the
space of all packings $P$ with density $1$, where ``density'' refers to the
usual concept of the limit of the proportion of the area of $P$ inside a
large disk centered at the origin as the radius tends to infinity.  Let $X'
\subset X$ be the space of
all packings $P$ such that $K_P$ is infinite and either the density of $P$
is less than
1, the density is not defined, or $t_P$ does not exist.
Thus, $X$ is the disjoint union of $X'$, $X''$ and 
the collection of $X_C$ for all $C$.

Let $\nu$ be any invariant probability measure with density $d$. Let $\nu_C$ be
$\nu$ conditioned on $X_C$, $\nu'$ be $\nu$ conditioned on $X'$ and $\nu''$ be
$\nu$ conditioned on $X''$. Since $\lambda_0$ is the only invariant probability
measure with support in $X''$, we have $\nu''=\lambda_0$. Thus,
$$
\nu = \nu(X')\nu' + \nu(X'')\lambda_0 + \sum_C \nu(X_C) \nu_C\,.
$$

Define the density $|\omega| := \omega(A_0)$ as before, but for any (invariant or non-invariant) probability
measure $\omega$ on $X$.
We have
$$
d=|\nu| = \nu(X')\,|\nu'| + \nu(X'') + \sum_C  \nu(X_C)\,|\nu_C|
$$
and
$$
\int f\,d\nu = \sum_C \nu(X_C) \int f\,d\nu_C = \sum_C \nu(X_C)
\frac{3}{j_C}|\nu_C|\,,
$$
where $j_C$ is the number of tiles in $C$.
\begin{lem}\label{lem5}
Let $\nu \in \tM$ and $C$ be a finite-component class.
Suppose that $0 \le  s \le 1$ is such that 
$|\cvxmu_s| = |\nu_C|$. Then
$\int f\,d\cvxmu_s \ge \int f\,d\nu_C$. Moreover, equality holds only if $j_C = 1$. 
\end{lem}

\begin{proof}
As in the proof of Lemma \ref{lem:cont}, we have
that $|\nu_C| = \int j_C/|V(t_P)| \,d\nu_C(P)$.

  First suppose that $j_C=1$.  Then $|\nu_C|
  = \int 1/|V(t_P)| \,d\nu_C(P) \le d_1$ by Lemma \ref{lem:lambda1}.
  This means that $s=1$ and $\int f \,d\nu_C
  =3\,|\nu_C|=3\,|\cvxmu_s|= \int f \,d\cvxmu_s$. 
  
  Now assume that $j=j_C > 1$ and put $p :={\rm \Per}_C$. By definition,
$$
\int f\,d\cvxmu_s = s \int f \,d\lambda_1 + (1-s) \int f
\,d\lambda_0= s \int f \,d\lambda_1 = 3sd_1\,.
$$
Since $\nu_C(\f)=3|\nu_C|/j_C = 3|\cvxmu_s|/j_C = 3(sd_1 + 1-s)/j$, it suffices to show that
$$
sd_1 > \frac{sd_1 + (1-s)}{j}\,,
$$
which is equivalent to
$$
s (j d_1 - d_1 + 1) > 1 \,.
$$
Now $sd_1 + (1-s) = |\nu_C| \le \frac{j}{ j + \ep p}$, where $\ep$ is from
Lemma \ref{lem:voronoi}. Solving for $s$ gives
$$
s \ge \frac{1- \frac{j}{j+\ep p}}{ 1-d_1}
\,,
$$
whence it is enough to show that
$$
(jd_1 -d_1 + 1)\frac{1- \frac{j}{ j+\ep p}}{ 1-d_1} > 1\,.
$$
This boils down to
$$
d_1(p\ep+1) > 1\,.
$$
Now, $j>1$ implies that $p \ge (7/6)\Per_1$, where $\Per_1$ is the perimeter
of a single tile. Since $\ep/100 > \delta$ (by Lemma \ref{lem:voronoi}), this
implies that $p\ep + 1 > 1 + \ep \Per_1 + \delta = 1/d_1$, proving the last
inequality.
\end{proof}

\begin{lem}\label{lem6} We have
$\int f\,d\nu \le \int f\,d\cvxmu_s$ 
for all $\nu\in {\widetilde M}$ with $|\nu|=|\cvxmu_s|$.
Equality holds only if
\begin{itemize}
\item $\nu(X_C)=0$ for every component class $C$ with $j_C > 1$ and
\item whenever $\nu(X_C) > 0$ and $j_C = 1$, we have $|\nu_C|=d_1$.
\end{itemize}
\end{lem}
\begin{proof}
Recall that
$$
\int f\,d\nu = \sum_C \nu(X_C) \int f\,d\nu_C\,.
$$
For each component class $C$, let $s_C$ be defined as follows:

\begin{itemize}
\item if there exists $s \in [0,1]$ such that $|\nu_C| = sd_1 + (1-s)$, then
set $s_C:=s$;
\item otherwise, set $s_C:=1$.
\end{itemize}
Let $\omega_C := s_C\lambda_1 + (1-s_C)\lambda_0$ and
$$
\sigma := \big(\nu(X')+\nu(X'')\big)\lambda_0 + \sum_C \nu(X_C)\omega_C 
\,.
$$
 From the previous lemma, if $|\nu_C| \ge d_1$, then $\int \f \,d\nu_C
\le \int \f \,d\omega_C$,
with equality only if $j_C = 1$. If $|\nu_C| < d_1$, then $s_C = 1$ and
$$
\int \f \,d\nu_C = \frac{3\,|\nu_C|}{j_C} < 3d_1 = \int \f \,d\omega_C\,.
$$

Summing up, we obtain
\begin{eqnarray*}
\int \f \,d\sigma &=& \sum_C\nu(X_C) \int f\,d\omega_C\\
&\ge& \sum_C\nu(X_C)\int f\,d\nu_C=\int f\,d\nu\,.
\end{eqnarray*}
Moreover, equality holds only if $\nu(X_C)=0$ for every component $C$
with $j_C > 1$ and $|\nu_C|=d_1$ whenever $j_C = 1$.
Since $|\omega_C| \ge |\nu_C|$, we have
\begin{eqnarray*}
|\sigma| &=& \nu(X')+\nu(X'') + \sum_C \nu(X_C)\, |\omega_C|\\
&\ge& \nu(X')\,|\nu'|+\nu(X'') + \sum_C \nu(X_C)\, |\nu_C|\\
&=&|\nu|\\
&=&|\cvxmu_s|\,.
\end{eqnarray*}
Since $\sigma$ and $\cvxmu_s$ are both convex combinations of $\lambda_0$
and $\lambda_1$, this 
implies that $\int f\,d\sigma\le \int f\,d\cvxmu_s$ with equality iff
$\sigma=\cvxmu_s$. Thus, $\int f\,d\nu\le \int f\,d\cvxmu_s$. 
In the equality case we must have $\int f\,d\nu = \int f\,d\sigma=\int
f\,d\cvxmu_s$ and
$\sigma=\cvxmu_s$. This 
implies that $\nu(X_C)=0$ if $j_C > 1$ and
$|\nu_C|=d_1$ 
if $j_C = 1$.
\end{proof}

\begin{lem}\label{lem3}
  Let $\nu \in \tM$.
  If $|\nu|=|\cvxmu_s|$, then $\int f\,d\nu
  \le \int f\,d\cvxmu_s$. Equality holds iff $\nu=\cvxmu_s$.
\end{lem}

\noindent
(Informally, $\cvxmu_s$ uniquely maximizes the number of degrees of freedom per tile
for invariant measures of a fixed density.)

\begin{proof}
  The previous lemma implies $\int f\,d\nu\le \int f\,d\cvxmu_s$. Assume
  $\int f\,d\nu = \int f\,d\cvxmu_s$; then
$$
\nu = \nu(X')\nu' + \nu(X'')\lambda_0 + \nu(X_C)\nu_C
$$
where $C$ is the component of size 1 and $|\nu_C|=d_1$. 
This gives $\int f\,d\nu = \nu(X_C)
3 d_1 = \int f\,d\cvxmu_s = 3s d_1$. Hence $\nu(X_C)=s$. Since $\nu'$ has density strictly
less than $1=|\lambda_0|$ but $|\nu|=|\cvxmu_s|$, we must have $\nu(X')=0$. That
is,
$$
\nu = \nu(X'')\lambda_0 + \nu(X_C)\nu_C
\,.
$$
Since $\nu$ and $\lambda_0$ are isometry invariant, $\nu_C$ must also be isometry 
invariant. By Lemma \ref{lem:lambda1}, $\lambda_1$ is the unique isometry-invariant
measure with support in $X_C$ and with density $d_1$. Hence $\nu_C = \lambda_1$. 
This implies $\nu=\cvxmu_s$ and the proof is finished.
\end{proof}

\medskip

\noindent
{\em Proof of Theorem~\ref{thm:main1}:}
It is easy to see that one can pack the $N\times N$ torus in such a way that there
is a large region of tight-linked tiles and a large region of densely loose-linked tiles,
in such a way that the interface between the two regions has a density which approaches
zero as $N$ tends to infinity, and the density $d_N$ of the packing $P_N$ tends to $d$.
Let $\omega_N$ be the invariant measure supported on isometric copies of ${\widetilde P_N}$
(a pull-back of $P_N$ to the plane). Then $\omega_N$ tends to $\cvxmu_s$ in the weak*
topology. By Lemma~\ref{lem:fcontinuity}, this implies that $\int
f\,d\omega_N \to \int f\,d\cvxmu_s$.
  
Now $\mu_N$, the uniform measure of density $d_N$ on the $N\times
N$ torus, satisfies $\int f\,d\mu_N \ge \int f\,d\omega_N$. This is because
$\mu_N$ is by definition supported on packings with the maximal
number of degrees of freedom for the given density $d_N$. Hence
$\liminf_N \int f\,d\mu_N \ge \liminf_N \int f\,d\omega_N = \int
f\,d\cvxmu_s$.
  
Therefore,
if $\mu_\infty$ is any weak* subsequential limit of $\Seq{\mu_N}_N$, then
$\int f\,d\mu_\infty \ge \int f\,d\cvxmu_s$. But $d_N \to d$, so $|\mu_\infty| =
|\cvxmu_s|$ by Lemma \ref{lem:cont}. The previous lemma now implies that
$\mu_\infty=\cvxmu_s$.  $\square$

Returning to the discussion of the introduction, we note that from
simulations of hard disks, one would expect the corollary to hold even
for a range of densities below $d_1$, but we do not know how to prove this.

\medskip

\noindent
{\bf Remark on Higher Dimensions}

\smallskip

The basic features of our argument can be generalized to dimension 3 or
higher, except for our use in Lemma \ref{lem:lambda1} of \cite{Fej2} on
the minimal Voronoi region in
disk packings in the plane. It would be of interest if this part of
our proof could be replaced by an argument insensitive to
dimension.

\section{Low Density}

In this final section, we confirm the intuition that at low densities,
there will be no infinite loose-linked component.  It is obvious that there
is no infinite {\em tight}-linked component at densities smaller than $d_1$.

We begin with
a lemma that holds for any tile shape (in fact, for any {\em collection}
of shapes and sizes, as long as each can be fit into a disk of some fixed
radius $s$, and ``density'' is interpreted as number of tiles per unit area).

\begin{lem}\label{lem:poisson} For small enough density $d$,
if a packing $P$ is drawn from $\limmu_d$, then the probability that the
disk $B_R$ of radius $R$ about the origin contains more than $9R^2d$ tile centers
goes to zero as $R \to \infty$.
\end{lem}

\begin{proof} Let $s$ be the radius of the smallest disk containing the tile
(in our case, $s$ is about $2^{1/2} \cdot 3^{-3/4} \cdot (1\!+\!2\rho)$) and choose
$$
0 < d < \frac{.05}{13 \pi s^2}~;
$$
for our zipper tiles with small enough $\rho$, $d \le .003$ suffices.
Let $T$ be the set of tiles whose centers fall in $B_R$, $k := \lceil \pi
R^2 d\rceil$
and $\ell > 9 R^2 d$.  Letting $\limmu_d(\cdot)$ denote
the probability of an event with respect to the measure $\limmu_d$,
we shall show that
$$
\frac{\limmu_d(|T|\!=\!\ell)}{\limmu_d(|T|\!=\!k)} \le \gamma^{\ell-k}
$$
for some constant $\gamma < 1$.  It then follows that
$$
\limmu_d(|T|> 9 R^2 d) \le {\limmu_d(|T|\!=\!k)}\sum_{\ell=\lceil 9R^2d \rceil}^\infty
\gamma^{\ell-k} \le
{\limmu_d(|T|\!=\!k)}\frac{\gamma^{\lfloor (9-\pi)R^2d \rfloor}}{1\!-\!\gamma}\to 0
$$
as $R \to \infty$, as desired.

The measure $\limmu_d$ is the limit of uniform distributions of
configurations on the $N \times N$ torus ${\bf T}_N$, in turn obtainable by
choosing a sequence of $n = \lfloor N^2 d \rfloor$ points from the
Lebesgue distribution $\lambda$ on ${\bf T}_N^n$ as the centers of the tiles,
orienting each tile independently and uniformly at random, and finally conditioning
on no overlap.  We denote by $\lambda(|T|\!=\!j)$ the
{\em a priori} probability that exactly $j$ points fall inside $B_R$ (which
we take to be some fixed disk in the torus).

Let $\Phi$ be the event that there is no overlap among the tiles whose centers
lie in $B_R$, and $\Psi$ the event that there is no overlap involving any tile
whose center falls {\em outside} $B_R$.  Then
$$
\frac{\limmu_d(|T|\!=\!\ell)}{\limmu_d(|T|\!=\!k)} = \frac{\lambda(|T|\!=\!\ell)}{\lambda(|T|\!=\!k)}
\cdot \frac{\lambda(\Phi\bigm||T|\!=\!\ell)}{\lambda(\Phi\bigm||T|\!=\!k)} \cdot
\frac{\lambda(\Psi\bigm||T|\!=\!\ell\,\wedge\,\Phi)}{\lambda(\Psi\bigm||T|\!=\!k\,\wedge\,\Phi)}~.
$$
and our job is to bound the three fractions on the right.

For the first, we note that $|T|$ is binomially distributed in the measure $\lambda$,
hence
\begin{eqnarray*}
\frac{\lambda(|T|\!=\!\ell)}{\lambda(|T|\!=\!k)} &=& \frac{ {n \choose \ell} \left(\frac{\pi R^2}{N^2} \right)^\ell
\left(1 - \frac{\pi R^2}{N^2} \right)^{n-\ell} }{ {n \choose k} \left(\frac{\pi R^2}{N^2} \right)^k
\left(1 - \frac{\pi R^2}{N^2}\right)^{n-k} }
\le\frac{(n\!-\!k)!/(n\!-\!\ell)!}{\ell!/k!} \left( \frac{
\frac{k}{n} }{ 1-\frac{k}{n} } \right)^{\ell-k}\\
&<& \frac{ (n\!-\!k)^{\ell-k} } { (\ell/e)^{\ell-k} } \cdot \frac { k^{\ell-k} } { (n\!-\!k)^{\ell-k} }
= \left(\frac{ek}{\ell} \right)^{\ell-k}
\le 0.95^{\ell-k}
\end{eqnarray*}
for large $R$.

The next fraction is easy:  since we may throw the first $k$ centers into $B_R$, then
the remaining $\ell\!-\!k$, we have
$$
\frac{\lambda(\Phi\bigm||T|\!=\!\ell)}{\lambda(\Phi\bigm||T|\!=\!k)}
$$
is the probability that the additional $\ell\!-\!k$ centers do not cause a collision, which is at most 1.

For the (inverse of) the third fraction, we throw $n\!-\!\ell$ centers into the region outside $B_R$,
then the remaining $\ell\!-\!k$.  A new point, if it lands at distance greater than $2s$ from
any previous point or from the disk $B_R$, causes no new overlap; and at each stage there
are fewer than $n\!-\!k$ points already placed.  Hence
$$
\frac{\lambda(\Psi\bigm||T|\!=\!k\,\wedge\,\Phi)}{\lambda(\Psi\bigm||T|\!=\!\ell\,\wedge\,\Phi)}
> \left( \frac{N^2 - \pi (R + 2s)^2 - (n\!-\!k)4 \pi s^2}{N^2 - \pi R^2} \right)^{\ell-k}
$$
$$
\ge \left(1 - 4 \pi s^2 d - \frac{4\pi sR + 4\pi s^2}{N^2 - \pi R^2}\right)^{\ell-k} 
> \left(1 - 13 s^2 d\right)^{\ell-k}
$$
for $N \gg R$.

Putting the inequalities together, we have
$$
\frac{\limmu_d(|T|\!=\!\ell)}{\limmu_d(|T|\!=\!k)}
\le \left(\frac{.95}{1-13 s^2 d}\right)^{\ell-k} = \gamma^{\ell-k}
$$
where $\gamma := .95/(1-13s^2 d) < 1$ by choice of $d$.
\end{proof}

\begin{pro}\label{thm:low} For some $d_2>0$, the $\limmu_d$-probability that the origin
is inside an infinite connected component of loosely-linked tiles is zero for $d<d_2$.
\end{pro}

\begin{proof}
Let $d \in (0,.003)$ be a density to be chosen later. Let $P$ be a
packing drawn from $\limmu_d$; we aim to show that the probability that
the origin is connected by a loose-linked chain of tiles of $P$ to
some point at distance $R$ approaches zero as $R \to \infty$.
  
We again choose some large radius $R$ and let $T$ be the set of tiles of $P$
whose centers fall inside the disk $B_R$. 

Fix the positions of the tiles of $P \setminus T$ (the black tiles of Figure~\ref{fig:chain})
and consider the space of packings having these tiles plus $n$ tiles whose centers
fall in $B_R$. We think of this space as being a subset of $T_1(B_R)^n/\Sym(n)$,
where $T_1(B_R)$ is the unit tangent bundle of $B_R$ (modulo a $2\pi/6$ rotation to
take into account the symmetries of the tile) and the symmetric group
acts by permuting the factors.

If $\alpha_n$ is the volume (in $T_1(B_R)^n/\Sym(n)$-space) of this
space and $m < n$, then by packing $n\!-\!m$ tiles into $B_R$ and then
the remaining $m$ in the left-over space, we have
$$
\alpha_n \ge \frac{1}{{n \choose {n-m}}}\alpha_{n-m} \frac{1}{m!}
\big[\pi(R-2s)^2 - n\pi(2s)^2\big]^m
\,,
$$
where $s$ is, as before, the radius of the circle circumscribing a tile.  This
takes into account possible intrusion of tiles in $P-T$ into $B_R$,
and the fact that a tile center at point $x$ can exclude nearby
centers but only within distance $2s$ of $x$.

\begin{figure}[htp]
\epsfxsize=3.5truein
$$\epsfbox{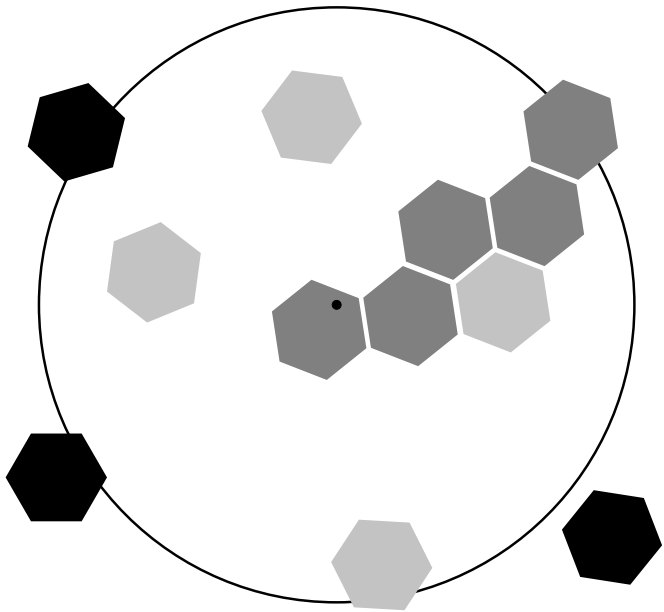}$$
\caption{An unlikely configuration of tiles in and around $B_R$}\label{fig:chain}
\end{figure}

Let $\beta$ denote the ``wiggle room'' of a tile $t$ loose linked to a
stationary tile $t'$, that is, the 3-dimensional volume of the space
of positions of $t$; then $\beta = \O(\rho^3)$ (but we use only
that $\beta$ is bounded by a constant).  If a packing ``percolates'',
that is, contains a chain of loose-linked tiles connecting the center
to the boundary of $B_R$, let $t_1, \dots, t_m$ be a shortest such
chain (the dark grey tiles of Figure~\ref{fig:chain}). Note that
$m \ge R/(2s)$.  For each $i>2$, the tile $t_i$ is linked to one of the
three sides of $t_{i-1}$ farthest from the side of $t_{i-1}$ linked to
$t_{i-2}$, and has wiggle room at most $\beta$ with respect to
$t_{i-1}$.  Accounting for the orientation of $t_1$ and allowing the
remaining $n\!-\!m$ tile centers to fall anywhere in $B_R$, we
have that the $3n$-dimensional volume of the set of percolating packings
is bounded by $(\pi/3)\cdot 6 \cdot 3^{m-2} \cdot \beta^{m-1} \cdot \alpha_{n-m}
< 3^m \beta^{m-1} \alpha_{n-m}$.

Comparing with the lower bound for $\alpha_n$, we find that given
$|T|=n \le 9dR^2$, the probability of percolation is less than
$$
\frac{3^m \beta^{m-1} \alpha_{n-m}}{ \alpha_n}
\le \frac{3^m \beta^{m-1} \, n!/(n\!-\!m)!}{\big[\pi(R-2s)^2 - n\pi(2s)^2\big]^m}
< \Big(\frac{27 \beta d R^2}{\pi\big[(R-2s)^2 - 36 d R^2 {s}^2\big]}\Big)^m /\beta
\,,
$$
which goes to zero as $R$ (thus also $m$) increases, for suitably chosen $d$.
Since we know from Lemma~\ref{lem:poisson} that $\limmu_d(|T| \le 9dR^2)$ approaches
1 as $R \to \infty$, the proposition follows.
\end{proof}

A more careful argument would prove Proposition~\ref{thm:low} for any density
below $1/\big(4\pi(2/3\sqrt{3})\big) = .2067^+$ for sufficiently small $\rho$, but clearly the probability of percolation
will remain 0 for much higher densities than that.

\section{A Conjecture}

We have shown that high-density random packings of zipper tiles in the plane contain
an infinite loose-linked component with positive probability, while
low-density random 
packings do not.
What happens in the case of ordinary disks, where there is no apparent linking
mechanism?  We believe, but cannot prove, the following

\medskip

\noindent
{\bf Conjecture.}  {\em Suppose $\limmu_d$ is defined as above
for geometric disks of radius 1.  Join two centers by an edge 
if their distance is at most $2 + \ep$ for some fixed $\ep \ll 1$. 
Then for sufficiently high density $d$ below the maximum, the graph
resulting from a configuration drawn from $\limmu_d$ will contain an
infinite connected component a.s.}

\smallskip
This statement can be shown by a standard Peierls-type argument for large
$\ep$; this may be known already.
In general, there is some parameter set of $(\ep, d) \subset (0, \infty)
\times (0, 1)$ for which there is an infinite component.
For small $d$ or for large $\ep$, the problem is quite similar to continuum
percolation, where one connects by an edge two points of a Poisson point
process if their distance is at most $r$.
Because of homotheties, one may fix the
intensity of the point process to be 1. Then there is a phase transition in
$r$. Our problem is quite different in really having two parameters,
but our conjecture is that there is a phase transition in $d$ for every
$\ep$ nevertheless.

\end{document}